\documentclass[hidelinks,preprint,3p,times,twocolumn]{elsarticle}

\usepackage{amssymb}
\usepackage{amsmath}

\usepackage{graphicx}

\journal{NIM-A}

\usepackage{orcidlink}
\usepackage{lmodern}
\usepackage{graphicx}
\usepackage{hyperref}

\usepackage[separate-uncertainty,exponent-product=\cdot,range-units=single,range-phrase=--]{siunitx}
\DeclareSIUnit{\neqpcm}{\si{n_{eq}\per\centi\meter\squared}}
\DeclareSIUnit{\electron}{e^-}
\DeclareSIUnit{\ppcm}{\si{p\per\centi\meter\squared}}
\DeclareSIUnit{\epcm}{\si{e^-\per\centi\meter\squared}}

\usepackage{xspace}
\newcommand{\BiOi}{$\text{B}_\text{i}\text{O}_\text{i}$\xspace}
\newcommand{\Bi}{$\text{B}_\text{i}$\xspace}
\newcommand{\Oi}{$\text{O}_\text{i}$\xspace}
\newcommand{\CiOi}{$\text{C}_\text{i}\text{O}_\text{i}$\xspace}
\newcommand{\Vtwo}{$\text{V}_2(+/0)$\xspace}
\newcommand{\Neff}{$N_\text{eff}$\xspace}
\newcommand{\Xdefect}{X-Defect\xspace}

\newcommand{\figuresref}[1]{\hyperref[#1]{\textcolor{black}{Figures~\ref{#1}}}}
\newcommand{\tablesref}[1]{\hyperref[#1]{\textcolor{black}{Tables~\ref{#1}}}}
\newcommand{\equationsref}[1]{\hyperref[#1]{\textcolor{black}{Equations~\ref{#1}}}}

\usepackage{makecell}
\usepackage{tabularx}
\usepackage{caption}

\usepackage{placeins}
\usepackage[T1]{fontenc}

\usepackage{microtype}

\usepackage{lineno}
\begin{document}

\begin{frontmatter}



\title{On the nature and charge state of the \Xdefect, a radiation-induced Silicon defect with field-enhanced charge carrier emission}

\author[a,b]{Niels G. Sorgenfrei\orcidlink{0000-0002-5729-6004}\corref{test}} 
\cortext[test]{Corresponding author}
\ead{niels.sorgenfrei@cern.ch}
\author[a]{Yana Gurimskaya\orcidlink{0000-0002-2549-4153}}
\author[a]{Anja Himmerlich\orcidlink{0000-0001-5418-0500}}
\author[a]{Michael Moll\orcidlink{0000-0001-7013-9751}}
\author[b]{Ulrich Parzefall}
\author[c]{Ioana Pintilie\orcidlink{0000-0002-3857-8524
}}
\author[d]{Joern Schwandt\orcidlink{0000-0002-0052-597X}}

\affiliation[a]{organization={CERN, European Organization for Nuclear Research},
            addressline={Esplanade des Particules 1}, 
            city={Geneva},
            postcode={1211}, 
            country={Switzerland}}
\affiliation[b]{organization={Institute of Physics, Albert-Ludwigs-Universitaet Freiburg},
            addressline={Hermann-Herder-Strasse 3}, 
            city={Freiburg im Breisgau},
            postcode={79104}, 
            country={Germany}}

\affiliation[c]{organization={National Institute of Materials Physics},
            addressline={Atomistilor 405A}, 
            city={Magurele},
            postcode={077125}, 
            country={Romania}}
\affiliation[d]{organization={Institute for Experimental Physics, University of Hamburg},
            addressline={Luruper Chaussee 149}, 
            city={Hamburg},
            postcode={22761}, 
            country={Germany}}

\begin{abstract}

The elusive \Xdefect, a defect found in low-resistivity $p$-type Silicon after irradiation, observed as a low-temperature shoulder of the \BiOi defect (Boron-interstitial-Oxygen-interstitial complex) in Thermally Stimulated Current (TSC) measurements, was investigated to determine its properties, matching them with those of a previously identified defect.
Through a combination of TSC, Deep-Level Transient Spectroscopy (DLTS), Difference-DLTS (DDLTS), numerical simulations of field-enhanced charge carrier emissions in TSC measurements and a comparison to literature, the \Xdefect was identified as the singly positively charged Silicon di-vacancy \Vtwo.
This assignment is supported by an agreement in activation energy, capture cross-section, trap type and charge emission process, as well as simulations comparing the effects of phonon-assisted tunnelling (PAT) and Poole-Frenkel (PF) mechanisms on TSC spectra.
DDTLS measurements revealed a quadratic dependence of the activation energy on the electric field strength, confirming PAT as the prevailing mechanism over PF in the case of the radiation-induced \Xdefect.
Assigning the \Xdefect to an electrically neutral defect in the space charge region resolves previous contradictions regarding its lack of impact on the effective doping concentration \Neff.

\end{abstract}

\begin{graphicalabstract}
\includegraphics[width=\textwidth]{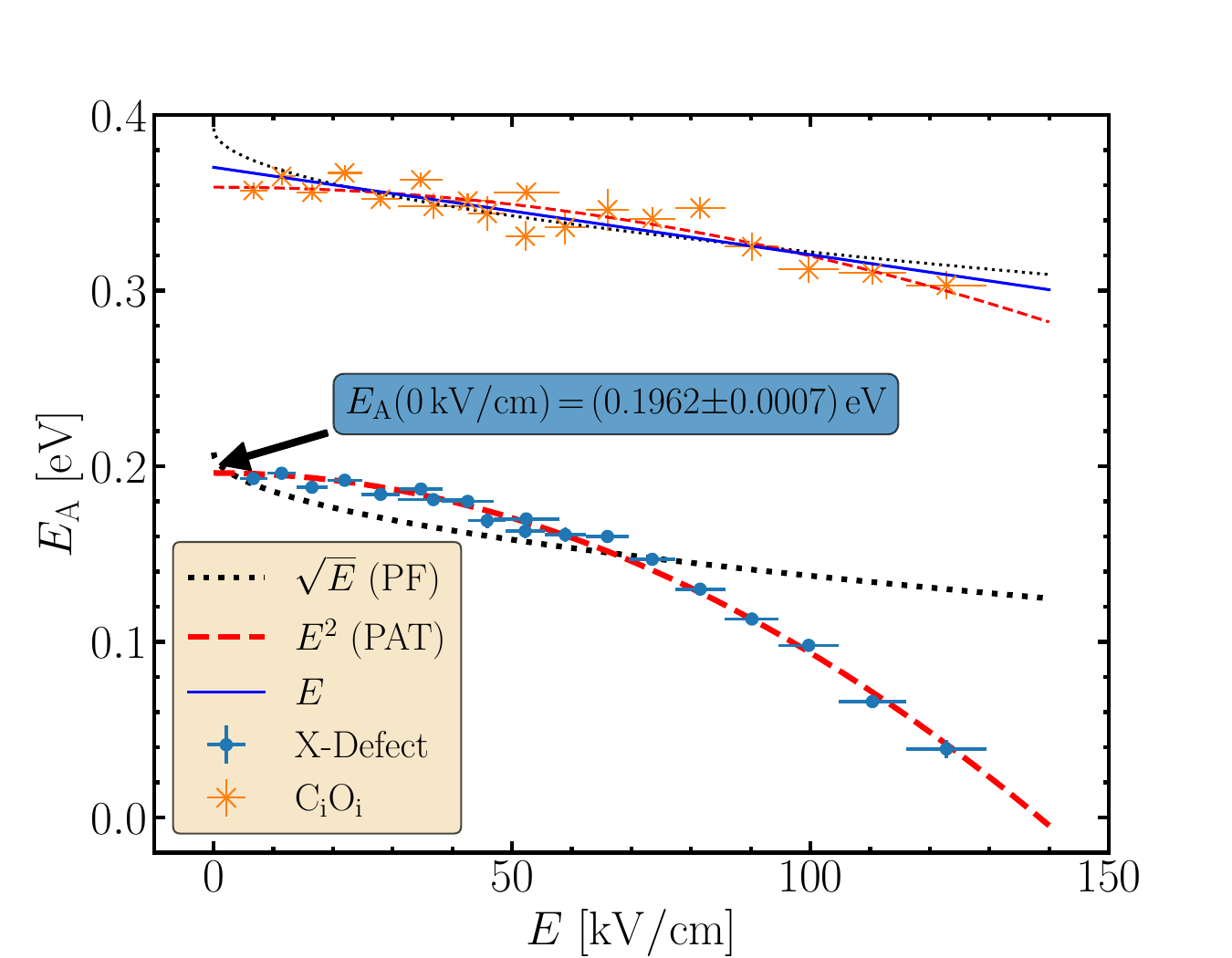}
This figure shows the electric field dependence of the activation energy for the \Xdefect. The red dashed line indicates a $E^2$ fit, which is the dependence expected for a phonon-assisted tunnelling process. This was the last missing property needed to match the \Xdefect's properties with those of the Silicon di-vacancy \Vtwo.
\end{graphicalabstract}

\begin{highlights}
\item Research highlight 1\\
\Xdefect is the Silicon di-vacancy in the donor state \Vtwo
\item Research highlight 2\\
\Xdefect does not impact \Neff
\end{highlights}


\end{frontmatter}



\section{Introduction \& History of the \Xdefect}
Radiation-induced defects in the Silicon lattice can significantly influence detector performance.
Energy levels from defects located deep within the bandgap result in increased leakage currents and enhanced carrier trapping, reducing the charge collection efficiency of a Silicon detector.
Examples of this are the $\text{V}_3$ (Silicon tri-vacancy) and the oxygen dependent $\text{I}_\text{P}$-centre point defects \cite{ioana1,ioana2,ioana3,ioana4}.
Deactivation of active dopant atoms, as well as the generation of Coulombic centres (ionised acceptors and donors) during irradiation introduces a change in the effective doping concentration \Neff.
Unlike usual trapping centres, such defects remain charged after ionisation at the sensor's operating temperature (usually around \SI{-20}{\celsius}), resulting in altered electric field strengths at a given bias voltage.
Key examples for the deactivation of dopants are the VP (Silicon-vacancy-Phosphorus-substitutional complex) and \BiOi (Boron-interstitial-Oxygen-interstitial complex) defects in $n$-type and $p$-type Silicon, respectively \cite{ioana5,ioana6,ioana7}.
Examples of Coulomb centres, defects with field enhanced emission, generated during irradiation, irrespective of the dopant type and with a strong influence on \Neff are the bi-stable donor (BD) centres, associated with thermal double donors (TDD2), and the acceptors H116K, H140K and H152K, associated with clusters of a small number of vacancies \cite{ioana10,ioana11,ioana12}.
Importantly, only defects charged in the depletion region at the usual operation temperatures can affect \Neff.

In irradiated low-resistivity $p$-type Silicon, the most detrimental effect on \Neff is due to the formation of \BiOi defects. 
This is known as the Acceptor-Removal Effect (ARE) and is currently considered to be the main contributor to the gain-layer degradation observed in Low Gain Avalanche Detectors (LGADs) \cite{michael_vertex}.
This defect is the result of a Boron dopant atom (acceptor) being knocked-out from its lattice site, becoming an interstitial (\Bi) and subsequently combining with an interstitial Oxygen (\Oi) impurity to form an electrically active defect, the \BiOi. 
This defect is a positively charged donor state in Silicon, effectively decreasing \Neff by double the amount of acceptors (Boron) removed when forming the \BiOi complex.
Various defect-engineering approaches, such as Phosphorus compensation or Carbon co-doping in LGADs, are being tested to improve gain-layer longevity \cite{valentina,ferrero}.

\begin{figure}[t]
    \centering
    \includegraphics[width=\linewidth]{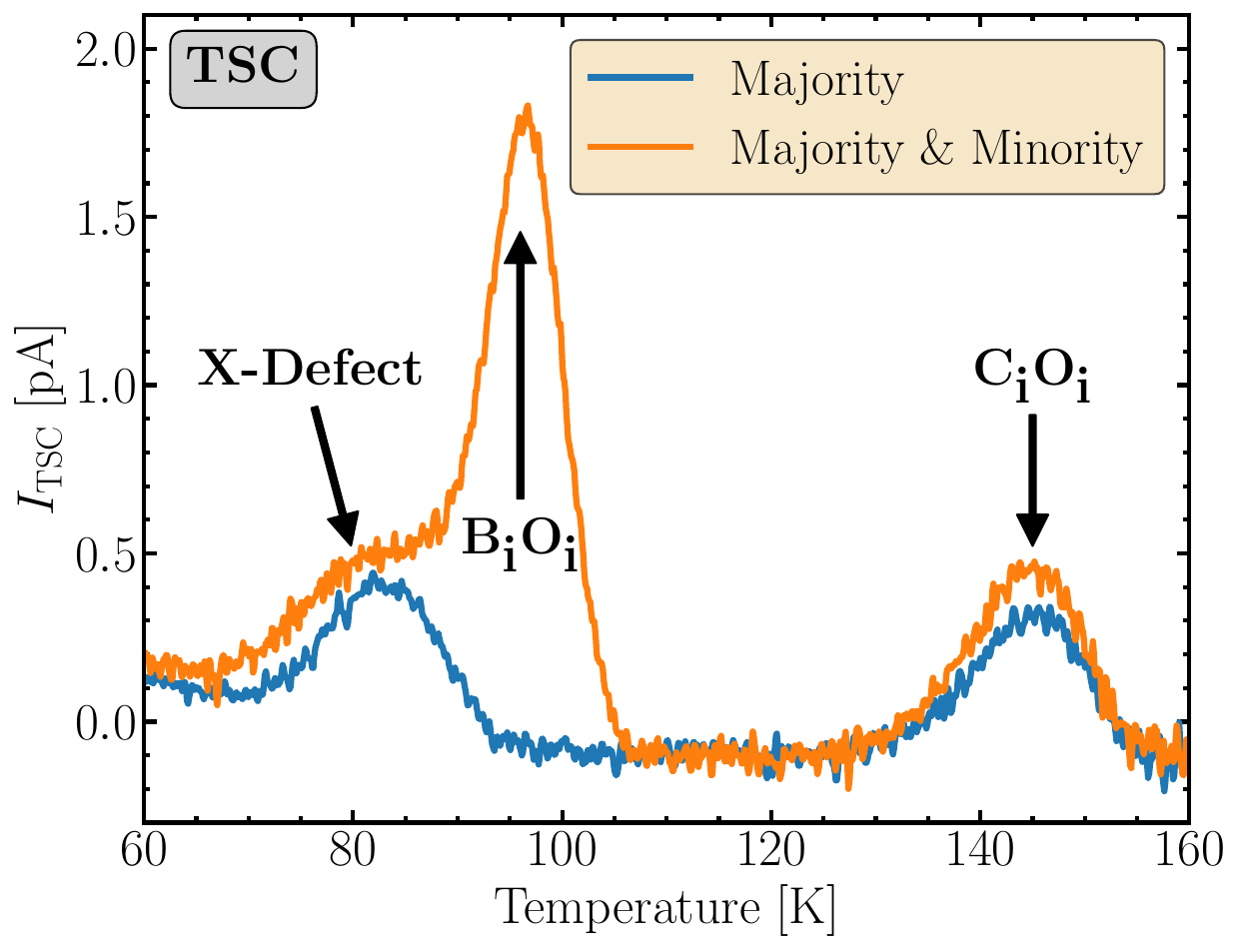}
    \caption{TSC measurements on the electron-irradiated diode. Measurement conditions: a reverse bias (during cooling and heating) of $U_\text{R}=\SI{20}{\volt}$, a filling temperature of $T_\text{fill}=\SI{60}{\kelvin}$ with either a \SI{0}{\volt} bias injection for majority carriers only (blue curve) or a \SI{-20}{\volt} forward bias (\SI{1}{\milli\ampere} current) injection for majority and minority carriers (orange curve), with a filling time of $t_\text{fill}=\SI{120}{\second}$ and a heating rate of \SI{11}{\kelvin\per\minute}. For the applied $U_\text{R}$ the peak electric field strength at the top of the diode is \SI{7,9}{\kilo\volt\per\centi\meter} and around \SI{5.2}{\micro\meter} are depleted at room temperature.}
    \label{fig:TSC_electron}
\end{figure}

In the context of investigations of the \BiOi defect in low-resistivity $p$-type Silicon, a presumably so-far not observed defect was discovered in Thermally Stimulated Current (TSC) measurements.
This defect appeared as a low-temperature shoulder to the \BiOi peak, see \autoref{fig:TSC_electron}, and showed an electric field enhanced emission and a temperature dependence of the carrier (holes) capture process \cite{liao1,anja1,liao2,liao3}.
No defect with such properties was previously reported as being induced by irradiation and it was therefore labelled the \Xdefect. 
Its emission rate, dependent on the electric field strength was tentatively explained by the Poole-Frenkel mechanism \cite{frenkel}.
This, however, would imply that the defect must be charged in the depleted region and therefore should impact \Neff at ambient temperatures.
This resulted in a puzzling observation: the change in concentration of the \Xdefect did not correlate with the change in the measured \Neff.

In this article, the \Xdefect is further investigated using Deep-Level Transient Spectroscopy (DLTS) and TSC measurements.
Electron- and $\gamma$-irradiated $p$-type Silicon diodes are studied, and simulations of charge carrier emission processes from defects are performed.
By comparing the measured characteristics of the \Xdefect with defect parameters reported upon in literature, a match of this defect with the singly charged di-vacancy in the donor charge state, \Vtwo, was found.

This article is a direct follow-up to our previous work Ref.~\cite{anja2}, in which the identification of the \Xdefect as the \Vtwo was already hypothetically proposed, but could not be verified.

In-depth explanations of the spectroscopic methods TSC and DLTS can be found in Ref.~\cite{Moll} and the references therein.
Since then, the TSC method was further developed both in terms of experimental procedures and methods to analyse the signal for obtaining accurate values for both defect parameters and defect concentrations.
Thus, efforts were made to reduce the large uncertainties in evaluating the TSC spectra, caused mainly by the variation of the space charge width during the carriers emission from traps, by the space charge sign inversion and by the possible asymmetry between the front and back electrode areas of the diodes. 
It has been shown that by measuring diodes with guard rings grounded and under reverse biases ensuring the full depletion of the diodes during the full temperature range, accurate analyses of the TSC signals can be performed \cite{ioana8}. 
Electric field enhanced emission processes, proving evidence for the charge states of detected defects can be nowadays accounted in numerical simulations of the TSC spectra \cite{ioana11, ioana9}.
Furthermore, for diodes that cannot be fully depleted over the TSC temperature scan (e.g. in highly doped diodes) precise evaluations of defect properties and concentrations can as well be made by performing combined thermally stimulated techniques for current (TSC) and capacitance (TSCap) experiments \cite{liao2}.
Only with these developments it became possible to establish a connection between microscopic and macroscopic properties in irradiated Silicon detectors. 
The method of the Difference-DLTS measurements will be briefly explained in \autoref{sec:ddlts} and is illustrated in Ref.~\cite{niels_3rdDRD3}.

\section{Experimental Methods and Materials}
Two $p$-type epitaxial Silicon pad diodes produced by \textit{CiS Forschungsinstitut für Mikrosensorik GmbH} \cite{cis} were investigated. The two diodes differ in their bulk resistivity but are otherwise identical.  
Both diodes have a thickness of \SI{50}{\micro\meter} and an active pad area (encased by the guard-ring structure) of \SI{6.927}{\milli\meter\squared}.
The guard-ring was grounded for all measurements presented.
The diode with a nominal resistivity of \SI{10}{\ohm\centi\meter} was irradiated with \SI{5.5}{\mega\electronvolt} electrons to a fluence of \SI{5e14}{\epcm} and has an effective doping concentration $N_\text{eff}$ of \SI{9.9e14}{\per\cubic\centi\meter} after irradiation.
The irradiation was carried out at the Belarusian state university in Minsk \cite{bsu}.
The other diode, with a nominal resistivity of \SI{250}{\ohm\centi\meter}, was irradiated with $^{60}$Co-$\gamma$ to a dose of \SI{2}{\mega\gray} and has a $N_\text{eff}$ of \SI{1.3E13}{\per\centi\meter\cubed} after irradiation.
The irradiation was carried out at the Ru\dj er Bo\v{s}kovi\'{c} institute in Croatia \cite{ruder}.

The low-resistivity (\SI{10}{\ohm\centi\meter}) sample was chosen for most of the measurements, as its low resistivity results in stronger electric fields for the same bias voltage, which becomes relevant for the electric field strength dependency measurements presented in \autoref{sec:ddlts}.
It has to be noted that the energy of the electrons used for the irradiation is slightly above the threshold value for producing clustered defects in Silicon \cite{jörn}.
This means that not only point-like defects, but also clustered defects can be created, which may distort peak shapes and activation energies.
However, only a minor influence from this effect is expected \cite{jörn}.
In the $\gamma$-irradiated sample, only point-like defects are expected.
This sample however, has a higher resistivity, resulting in lower electric field strengths.
Ideally, low-resistivity $\gamma$-irradiated samples would be best suited for this study, but were not available.

The DLTS measurements were performed using the commercial system FT 1030 from PhysTech \cite{Phystech}.
The TSC measurements utilised a Keithley 6517A and a custom LabVIEW software.
A closed-cycle liquid Helium cryocooler from ARS \cite{ars} was used to control the temperature of the samples from \SI{20}{\kelvin} to \SI{350}{\kelvin}.
\section{Results and Discussion}

\begin{figure}[t]
    \centering
    \includegraphics[width=\linewidth]{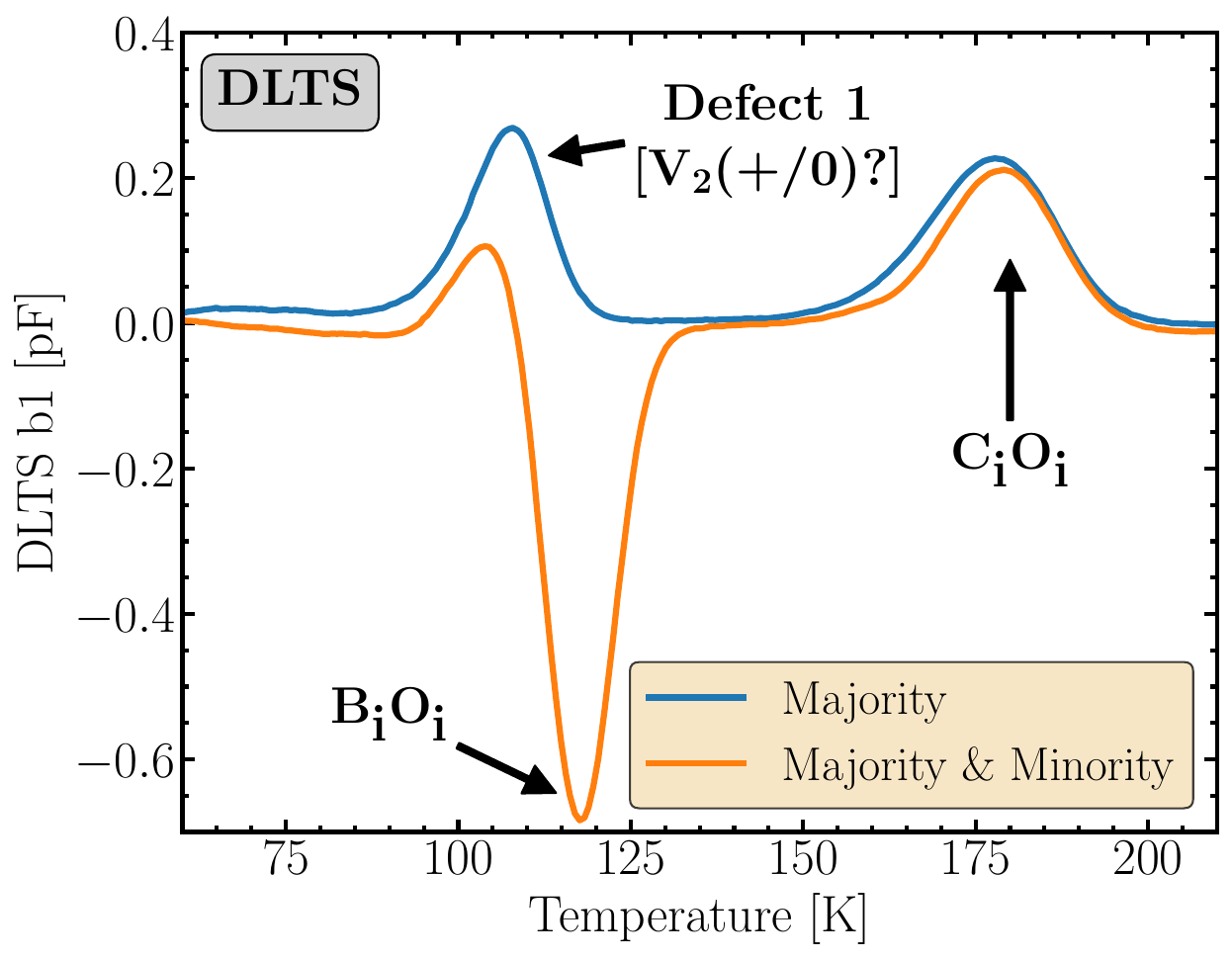}
    \caption{DLTS measurements for the electron-irradiated diode. Measurement conditions: a reverse bias of $U_\text{R}=\SI{10}{\volt}$, pulse voltages of $U_\text{P}=\SI{0.6}{\volt}$ reverse bias for a majority carrier only filling (blue curve) and \SI{-2}{\volt} forward bias for a majority and minority carrier filling (orange curve), with a pulse duration of $t_\text{p}=\SI{1}{\milli\second}$, evaluated for time windows $T_\text{W}$ of \SI{20}{\milli\second}, \SI{200}{\milli\second} and \SI{2}{\second}. Shown are the spectra obtained from the b1 correlator and $T_\text{W}=\SI{200}{\milli\second}$. For the applied $U_\text{R}$ the peak electric field strength at the top of the diode is \SI{5,7}{\kilo\volt\per\centi\meter} and around \SI{3.8}{\micro\meter} are depleted at room temperature.}
    \label{fig:DLTS_electron}
\end{figure}
Figures \ref{fig:TSC_electron} and \ref{fig:DLTS_electron} show the results from TSC and DLTS measurements on the electron-irradiated diode.
\autoref{fig:TSC_electron} shows the \Xdefect as a left shoulder to the \BiOi defect peak.
By comparing the two filling conditions (only majority in blue, and both majority and minority charge carrier filling in orange), it can be inferred that the \Xdefect is a hole trap.
Since the \BiOi is an electron trap, the \Xdefect can be studied without interference from the \BiOi peak using a majority carrier only filling measurement (blue line).

The DLTS spectra in \autoref{fig:DLTS_electron} obtained on the same diode show similar results.
The \BiOi has a left shoulder (on the orange curve), labelled as Defect 1.
In the following sections, it is shown that Defect 1 corresponds to the \Xdefect, and its properties resemble those of the single charged donor level of the di-vacancy \Vtwo.

\subsection{Identification of \Xdefect in TSC and DLTS}
Both, the DLTS and the TSC measurements, identify Defect 1 and the \Xdefect as hole traps.
Both measurements equally show the \BiOi and \CiOi defects, as well as one additional defect appearing as a low-temperature shoulder to the \BiOi signal (TSC) or a defect distorting the \BiOi peak towards its lower temperature tail (DLTS).
However, this alone does not suffice to conclude that the \Xdefect observed in TSC equals Defect 1, even though the extracted defect parameters are close. 
It is not straightforward to compare TSC and DLTS spectra, since one is based on current measurements  and the other one on capacitance measurements. 
Furthermore, the temperature at which a defect peak appears is shifted between the two measurement methods and depends strongly on the measurement conditions (e.g. the rate window for DLTS and the heating rate for TSC).
Therefore, additional effort is needed to match the results obtained by the two methods. 
In this study, a simulation of what a TSC spectrum would look like was performed using defect parameters extracted from DLTS measurements to enable a more direct comparison between the two methods. 
More details about the custom simulation framework is given in \autoref{sec:simulation}.

\begin{figure}[t]
    \centering
    \includegraphics[width=\linewidth]{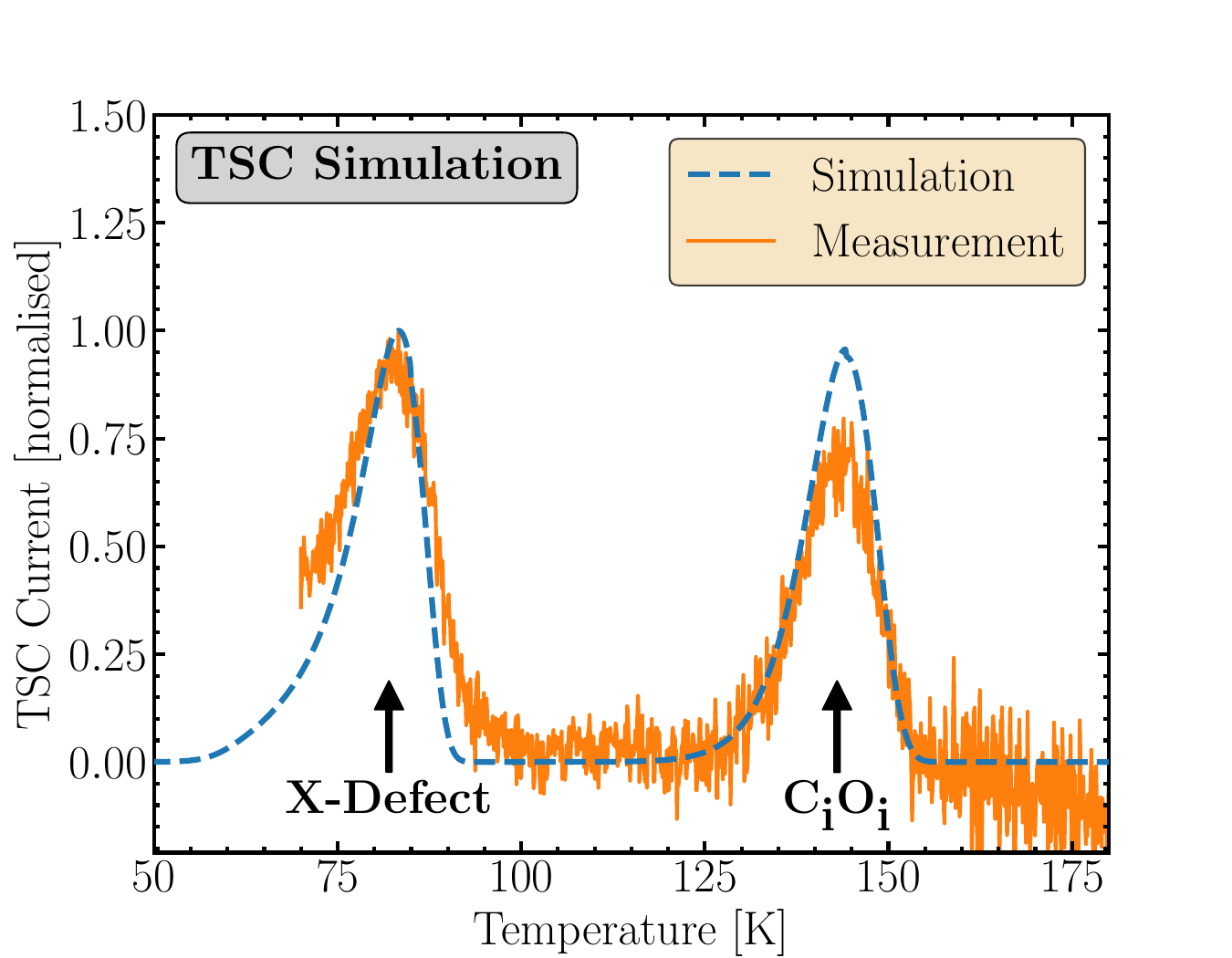}
    \caption{Simulation of a TSC spectrum using defect parameters obtained from DLTS measurements in \autoref{fig:DLTS_electron} (blue curve). The TSC measurement was obtained for a reverse bias of \SI{20}{\volt}, a filling temperature of \SI{70}{\kelvin} and a filling voltage of \SI{0}{\volt} applied for \SI{120}{\second}.}
    \label{fig:simulation_vs_measurement_electrons}
\end{figure}

\begin{figure}[t]
    \centering
    \includegraphics[width=\linewidth]{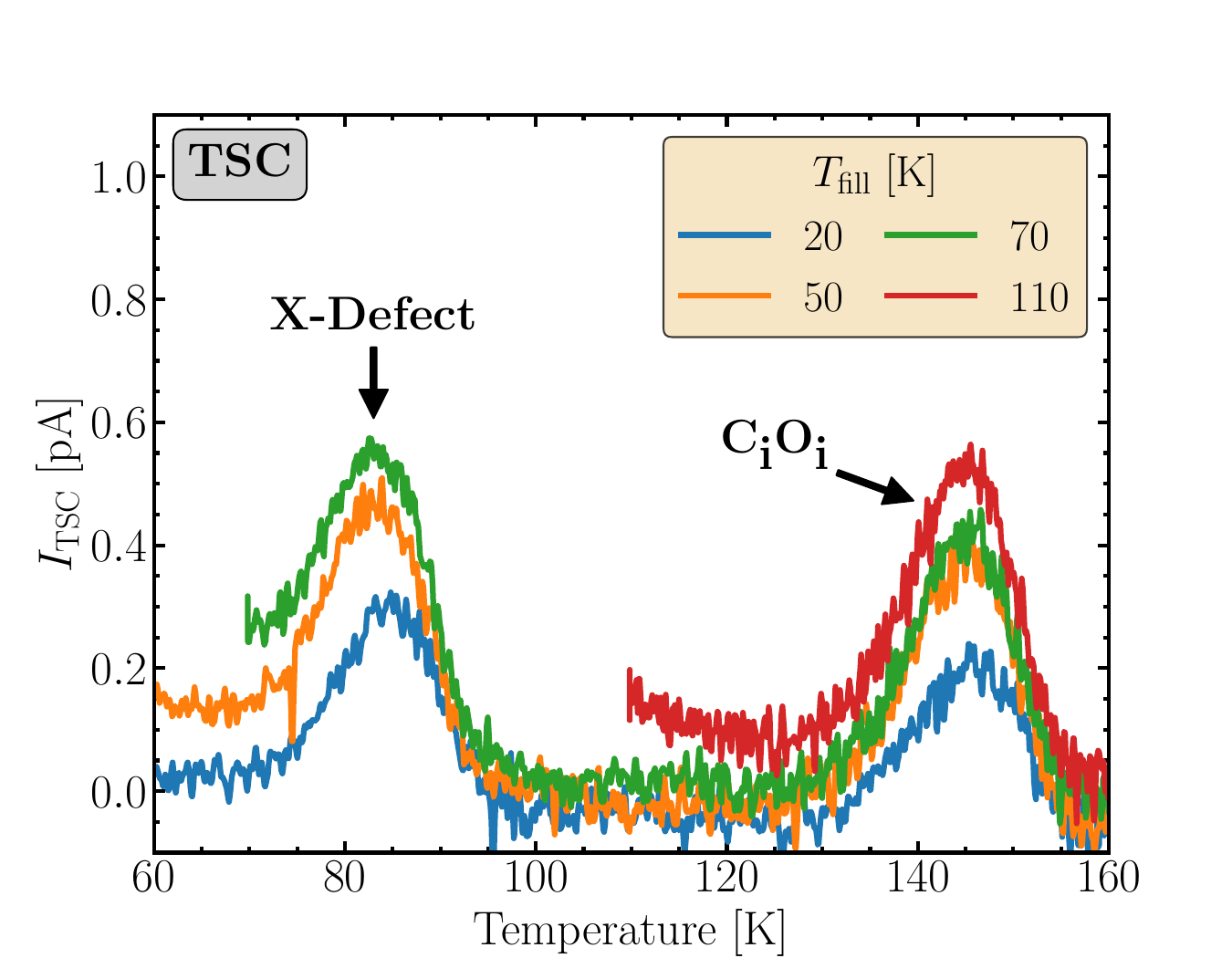}
    \caption{Effect of the filing temperature $T_\text{fill}$ in TSC measurements. Obtained for the electron-irradiated sample with a reverse bias of \SI{20}{\volt} and a filling voltage of \SI{0}{\volt} for \SI{120}{\second}.}
    \label{fig:TSC_Tfill}
\end{figure}

The result of these simulations, superimposed with a TSC measurement, is shown in \autoref{fig:simulation_vs_measurement_electrons}.
The TSC measurement conditions were chosen such that a large fraction of the \Xdefect-related traps in the diode were filled.
Both the \Xdefect and \CiOi exhibit a strong dependence on the filling temperature $T_\text{fill}$.
This is observed in \autoref{fig:TSC_Tfill}, where under otherwise identical measurement conditions, the electron-irradiated sample was repeatedly measured by using the TSC method, while only varying the filling temperature.
Clearly, at higher $T_\text{fill}$, a larger fraction of the defect-related traps are filled, resulting in higher peak amplitudes.

In \autoref{fig:simulation_vs_measurement_electrons} it can be seen, that the positions and shapes of both measured peaks, the \Xdefect and \CiOi, are reproduced by the simulated spectrum.
Some discrepancies between the measured and simulated spectra are visible.
For one, the absolute currents do not match with the measured ones.
The DLTS method measured concentrations about one order of magnitude lower than the TSC method.
Therefore, the signal normalised to the \Xdefect's amplitude is displayed in the figure.
This discrepancy between the two methods could not be explained.
Furthermore, the deviation of the amplitude of the \CiOi peak, even after the normalisation to the X-Defect's peak, arises from its dependence on the filling temperature and the different filling procedures specific to the DLTS and TSC methods.
While in DLTS, filling pulses are applied for each temperature in the temperature ramp, in TSC the filling is done once, at the lowest temperature.
Therefore, for centres with temperature dependent capture cross sections like \CiOi, the TSC peak will rise as the filling temperature increases.
Looking at the spectrum for a filling temperature of \SI{110}{\kelvin} in \autoref{fig:TSC_Tfill}, it is clear that the filling of \CiOi is not yet saturated at $T_\text{fill}=\SI{70}{\kelvin}$.

Most of the observed differences arise from highly idealised assumptions in the simulation of charge carrier emission in $n$-$p$ junctions, e.g. constant doping profiles, idealised triangular electric field shapes, no recombination or trapping effects and no charge carrier freeze-out, to name a few.
Nonetheless, the simulation gives strong confidence to conclude that both signals, from Defect 1 detected in DLTS and from the \Xdefect detected in TSC, belong to the same defect.

\subsection{Assignment of \Xdefect to \texorpdfstring{V$_\mathit{2}$(+/0)}{V2(+/0)}}
The next step in this study is to demonstrate, that the properties of the \Xdefect, now considered synonymous to Defect 1, match those of an already identified radiation-induced defect. 
Among the multitude of energy levels detected in irradiated Silicon, the donor-charge state of the di-vacancy, \Vtwo, has the closest matching trapping parameters to those of the \Xdefect. 
The defect parameters of the \Xdefect and the \Vtwo, as reported by Zangenberg et al. \cite{zangenberg}, are listed in \autoref{tab:matchmaker}.

\begin{table}[h]
\centering
\resizebox{\linewidth}{!}{%
\begin{tabular}{c|cc}
    & \makecell{This work\\ \Xdefect} & \makecell{Literature \cite{zangenberg}\\ \Vtwo} \\\hline
Energy Level $E_\text{A}$ [\si{\electronvolt}]        & \num{0.186} & \num[separate-uncertainty=false]{0.189(3)} \\
\makecell{Capture Cross\\Section $\sigma$ [\si{\per\centi\meter\squared}]} & \num{2.98e-16} & \num{2.59e-16} \\
Type of Defect & hole & hole \\
\makecell{E-Field Dependent\\Emission Process} & \makecell{phonon-assisted\\tunnelling} & \makecell{phonon-assisted\\tunnelling}
\end{tabular}
}
\caption{Comparison of literature values/characteristics for the \Vtwo from Ref.~\cite{zangenberg} and DLTS measurement results for Defect 1.}
\label{tab:matchmaker}
\end{table}

An exact comparison of the energy level value is difficult, due to the defect's electric field dependent emission process, effectively making $E_\text{A}$ dependent on measurement conditions and the material used.
The value shown in \autoref{tab:matchmaker} is for the electron-irradiated sample, a \SI{10}{\ohm\centi\meter} $p$-type diode, from DLTS measurements with a reverse bias of \SI{10}{\volt} and a (reverse) pulse voltage of \SI{0.6}{\volt} for a majority carrier filling.
The reference article \cite{zangenberg} to which the measurement results are compared to, reports their DLTS measurement conditions as $U_\text{R}=\SI{5}{\volt}$ and $U_\text{P}=\SI{1}{\volt}$, but does not specify the effective doping concentration of their sample.
Other studies cite values of \SI{0.191}{\electronvolt} \cite{zangenberg2} and \SI{0.194}{\electronvolt} \cite{hallen}.
Due to the defect's electric field-dependent emission process, the extrapolated zero-field activation energy $E_\text{A,0}$ should be quoted.
Nonetheless, a good qualitative match of the activation energies is found.
Furthermore, the capture-cross sections differ only slightly, and the differences are well within the measurement uncertainty of \SI{25}{\percent} \cite{niels_4HSiC}.
As \Vtwo, the \Xdefect also acts as a hole trap, as shown in \autoref{fig:DLTS_electron}, where the \Xdefect is visible for the majority carrier only filling measurement conditions.

These findings were also presented in earlier works, most recently in Ref.~\cite{anja2}.
The final missing information to conclusively match the \Xdefect with \Vtwo is the field-dependent emission process.
Up to now, it was assumed that the \Xdefect follows a Poole-Frenkel mechanism \cite{liao1}.
However, this was never verified and only assumed, since no other emission processes were considered at that time.
In the following section, the two relevant emission processes and the required measurements to distinguish them are described.

\subsubsection{E-Field Dependence of the Emission Rate}\label{sec:ddlts}
\begin{figure}[t]
    \centering
    \includegraphics[width=\linewidth]{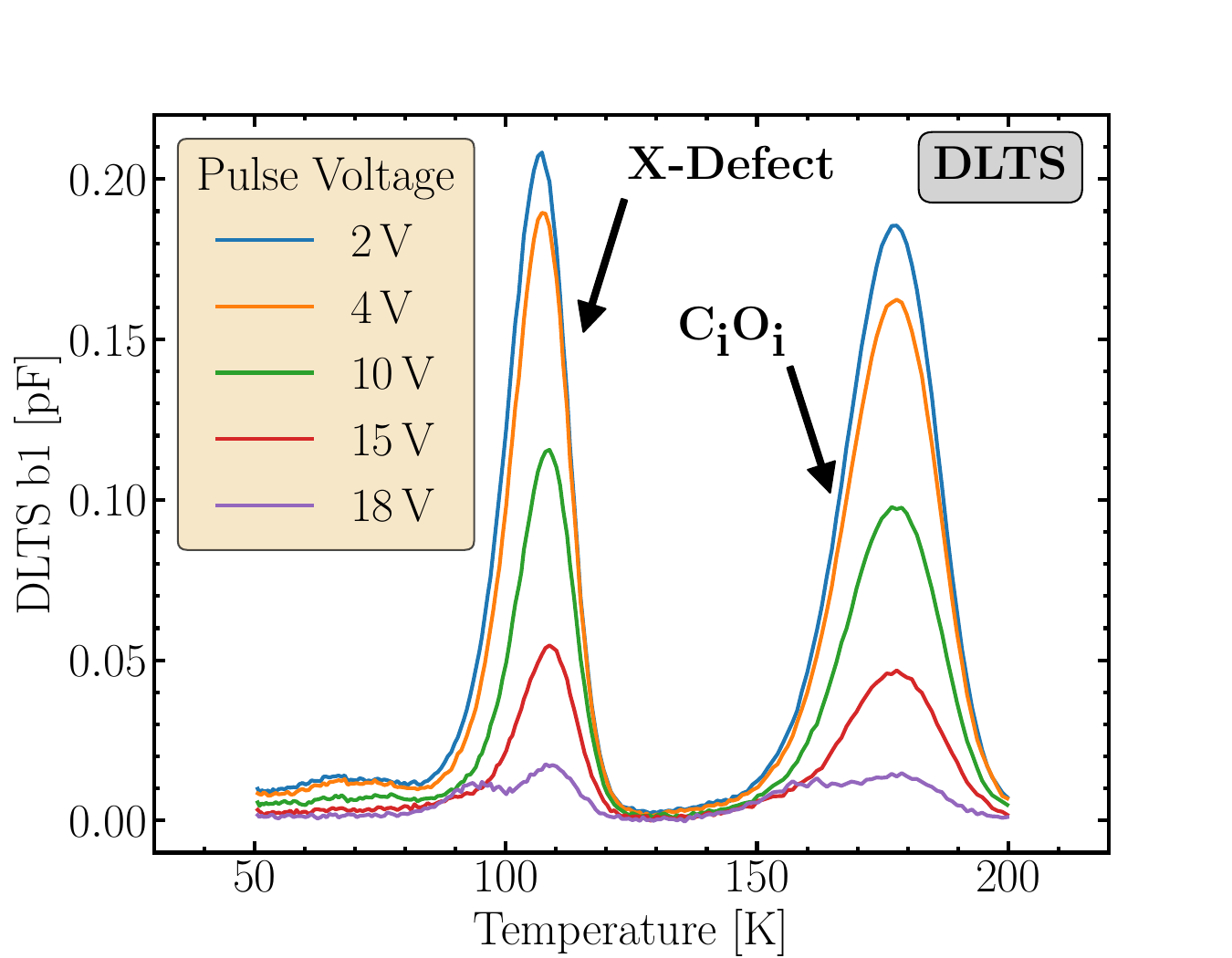}
    \caption{DLTS measurements for different pulse voltages. All measurements were carried out under a reverse bias of \SI{20}{\volt}.}
    \label{fig:DLTS_pulse}
\end{figure}
\begin{figure}[t]
    \centering
    \includegraphics[width=\linewidth]{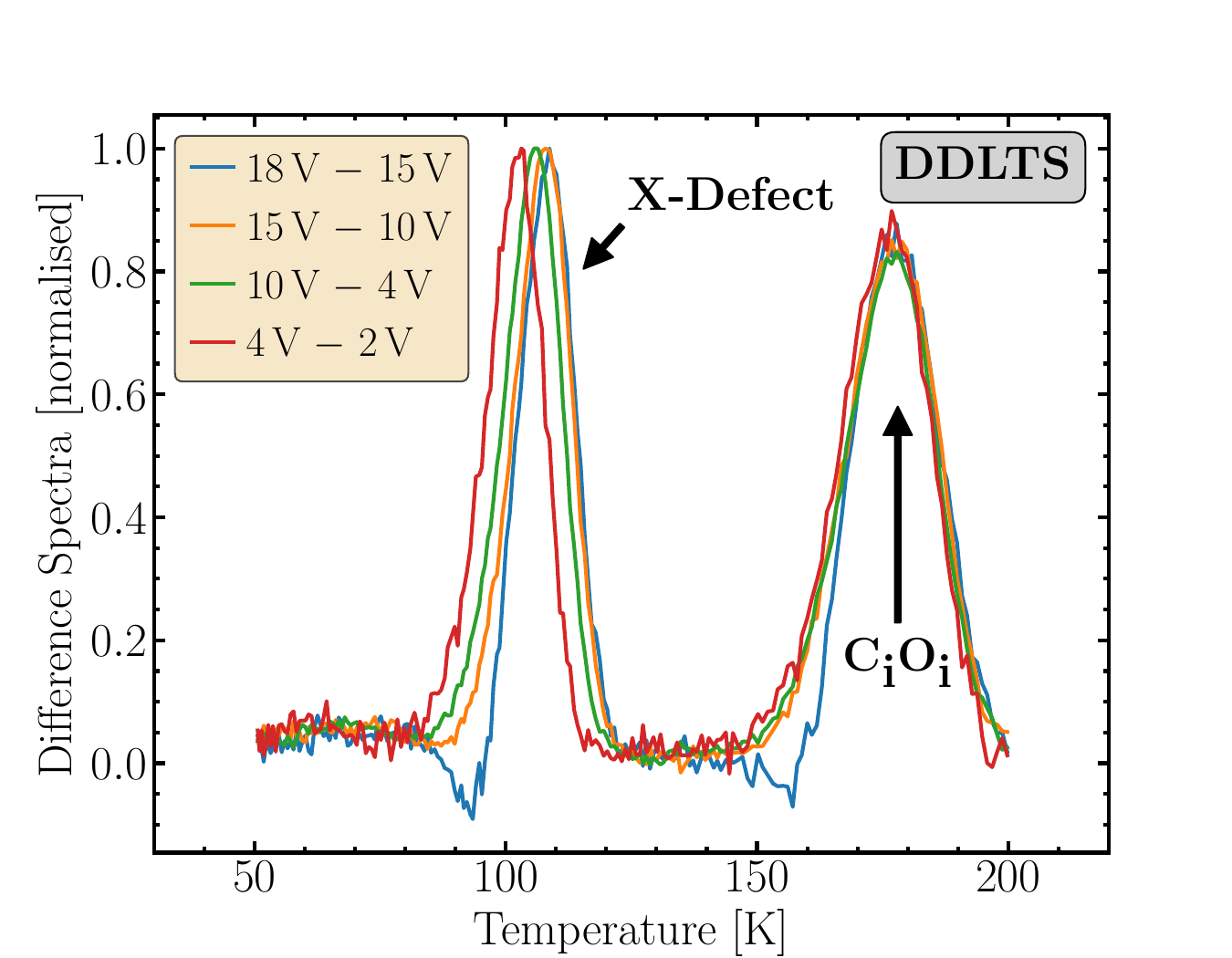}
    \caption{DLTS difference spectra obtained from subtracting the measurements shown in \autoref{fig:DLTS_pulse} from each other. This results in effectively observing only signals induced by charge carriers under certain electric field strengths.}
    \label{fig:DLTS_difference}
\end{figure}
\begin{figure}[t]
    \centering
    \includegraphics[width=\linewidth]{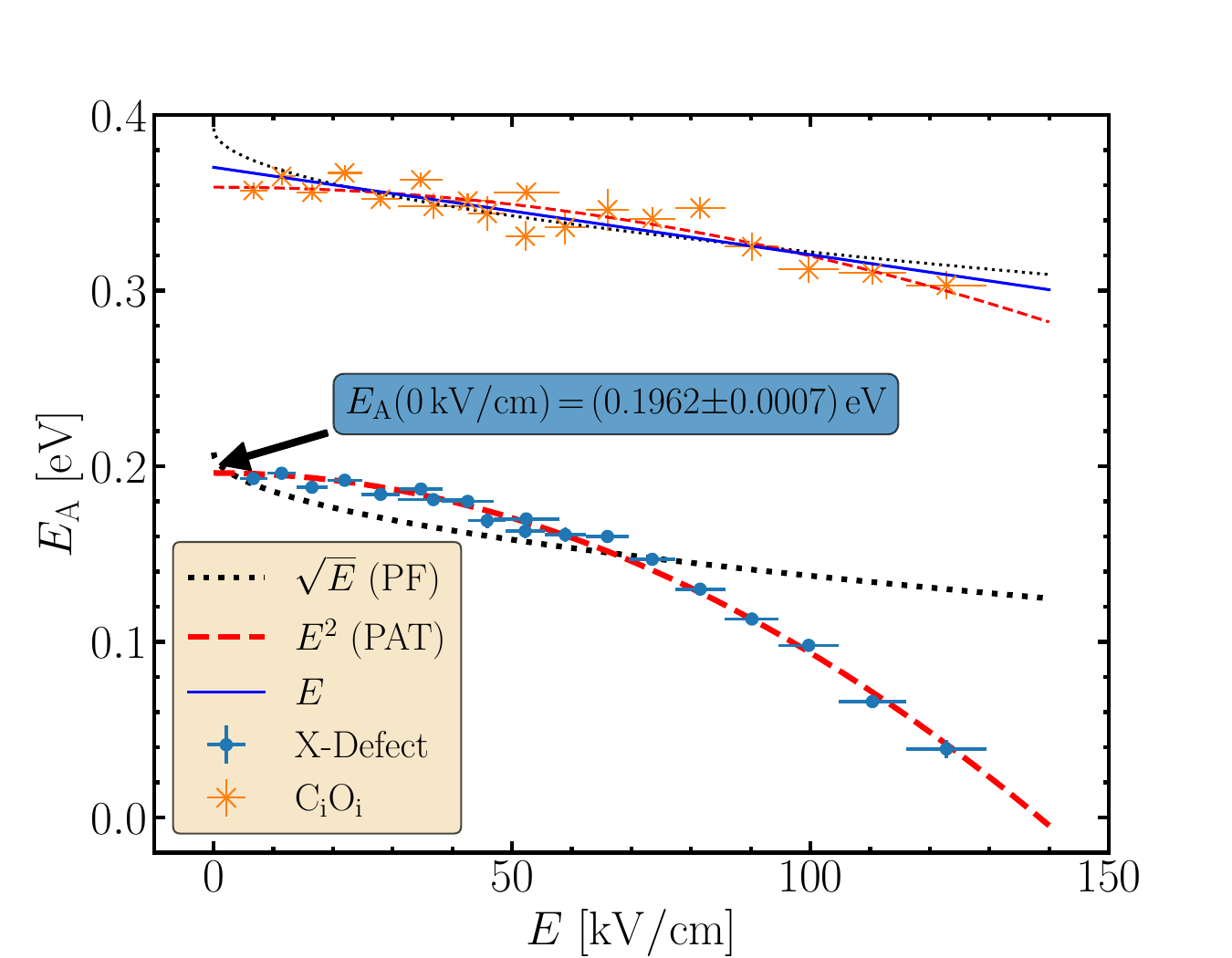}
    \caption{Dependence of the activation energy $E_\text{A}$ on the electric field strength $E$. The results for the \Xdefect are shown in light blue points, and orange for the \CiOi. Functions of the form $\sqrt{E}$ for the PF (dotted black line), $E^2$ for the PAT mechanism (dashed red line) and $E$ linearly (solid blue line, only for \CiOi) are fitted onto the datapoints.}
    \label{fig:DLTS_efield_dependence}
\end{figure}

Defects can have an electric field strength dependent enhanced emission.
This is seen in TSC measurements as a shift of the peak maxima to lower temperatures with increasing applied biases.
Not all defects are affected by this, and there are different processes which can cause this behaviour.
The two mechanisms relevant for this study are the Poole-Frenkel (PF) effect and phonon-assisted tunnelling (PAT) mechanism.
The PF is only applicable for Coulombic centres, defects that are charged after the emission of the charge carrier.
This means that the ionised defect and the emitted charge carrier interact via the Coulomb interaction and that the applied voltage can lower the potential barrier for the charge carrier emission. 
Examples of such centres are the Phosphorus and Boron dopants and the \BiOi defect. 
Neutral defects can, however, have an enhanced emission via the PAT mechanism.
One way to differentiate between the two processes, when one does not know the charge state of the studied defect, is a method suggested first by Ganichev et al. \cite{ganichev}. 
PF and PAT differ in the way the charge carrier emission from a defect is enhanced by an electric field strength $E$.
While the PF shows a $\sqrt{E}$ dependence of the activation energy, PAT shows an enhanced emission following a $E^2$ dependence.
The activation energy is a precisely measurable quantity and therefore, this method yields a clearly identifiable distinction between the two mechanisms.

To obtain the zero-field activation energy value $E_\text{A,0}$ and determine whether the \Xdefect exhibits a phonon-assisted tunnelling or Poole-Frenkel enhanced emission, Difference-DLTS (DDLTS) measurements were carried out.
The DDLTS method works by subtracting two DLTS measurements on a transient-by-transient basis from one another, where they only differ by the applied pulse voltage.
Here, pulse voltages only in the reverse bias direction were used, since this enables a study of the \Xdefect without interference from the \BiOi defect.
Different pulse voltages in a diode result in a different depth range of filled defects.
By subtracting these measurements from one another, only defects in a small depth range are contributing to the observed signal.
This, however, also means that they experience only a narrow slice of the full electric field profile in the diode during the emission step.\footnote{This method is instructively illustrated in Ref.~\cite{niels_3rdDRD3}.}

The electric field profile is calculated by assuming an abrupt junction in a planar pad diode \cite{sze}.
If the applied voltage is smaller than the full depletion voltage $V_\text{FD}$, the electric field as a function of depth $x$ is given by the following expression, depending on whether $x$ is smaller or larger than the depletion depth $w$:
\begin{align}
E(x) = \begin{cases}
\dfrac{q N_\text{eff}}{\varepsilon_0\varepsilon_\text{r}}(w - x), & x < w, \\
0, & x \ge w.
\end{cases}\label{eq:efield}
\end{align}
In the case of over-depletion, it is calculated as \cite{ioana9}
\begin{align}
E(x) = \dfrac{q N_\text{eff}}{\varepsilon_0\varepsilon_\text{r}}(w - x) + \dfrac{V_\text{tot}-V_\text{FD}}{d},\label{eq:efield_overdepletion}
\end{align}
with $\varepsilon_0$ being the vacuum permittivity, $\varepsilon_\text{r}$ the relative permittivity of Silicon and $d$ the device thickness.
The effective doping concentration \Neff and applied reverse bias $V_\text{bias}$ determine the depletion depth $w$ with
\begin{align}
w = \sqrt{\dfrac{2 \varepsilon V_\text{tot}}{q N_\text{eff}}},\label{eq:depletion_depth}
\end{align}
where $V_\text{tot}=V_\text{bias}+V_\text{bi}$, and $V_\text{bi}$ being the built-in voltage set to \SI{0.7}{\volt}.

Some exemplary DLTS measurements for different pulse voltages are shown in \autoref{fig:DLTS_pulse}.
As can be seen, a smaller pulse voltage results in a larger signal amplitude, due to a larger volume with defects being filled contributing to the signal generation.
\autoref{fig:DLTS_difference} shows examples of difference spectra obtained after subtracting the pulse measurements from one another, transient by transient.
Clear shifts in the positions for the \Xdefect peaks are observed, while the \CiOi peaks exhibit none, indicating that the \Xdefect is subject to a field dependent emission process, while the \CiOi is not.
For these spectra the standard DLTS analysis with correlator functions is performed to construct an Arrhenius plot, which yields the activation energy. 
The resulting activation energy versus the applied electric field strength is displayed in \autoref{fig:DLTS_efield_dependence}.
The measurement results show that the field strength dependence for the \Xdefect activation energy fits well to an $E^2$ dependence, the expected behaviour for phonon-assisted tunnelling.
An approach to fit a $\sqrt{E}$ dependence, i.e a PF related dependency, does not result in a good data representation.
The least-squares fitting of 
\begin{align}
    f(E)=a\cdot E^2+b \label{eq:E2}    
\end{align}
to the datapoints results in a zero-field activation energy of $E_\text{A,0}=b=\SI{0.1962(7)}{\electronvolt}$.
The slope parameter for the fit is $a=\SI{-1.024(14)e-5}{\electronvolt\centi\meter\per\kilo\volt}$.
The reduced $\chi^2$ is \num{2,49}, indicating a good fit of the PAT $E^2$ model to the measured data.

\begin{figure}[t]
    \centering
    \includegraphics[width=\linewidth]{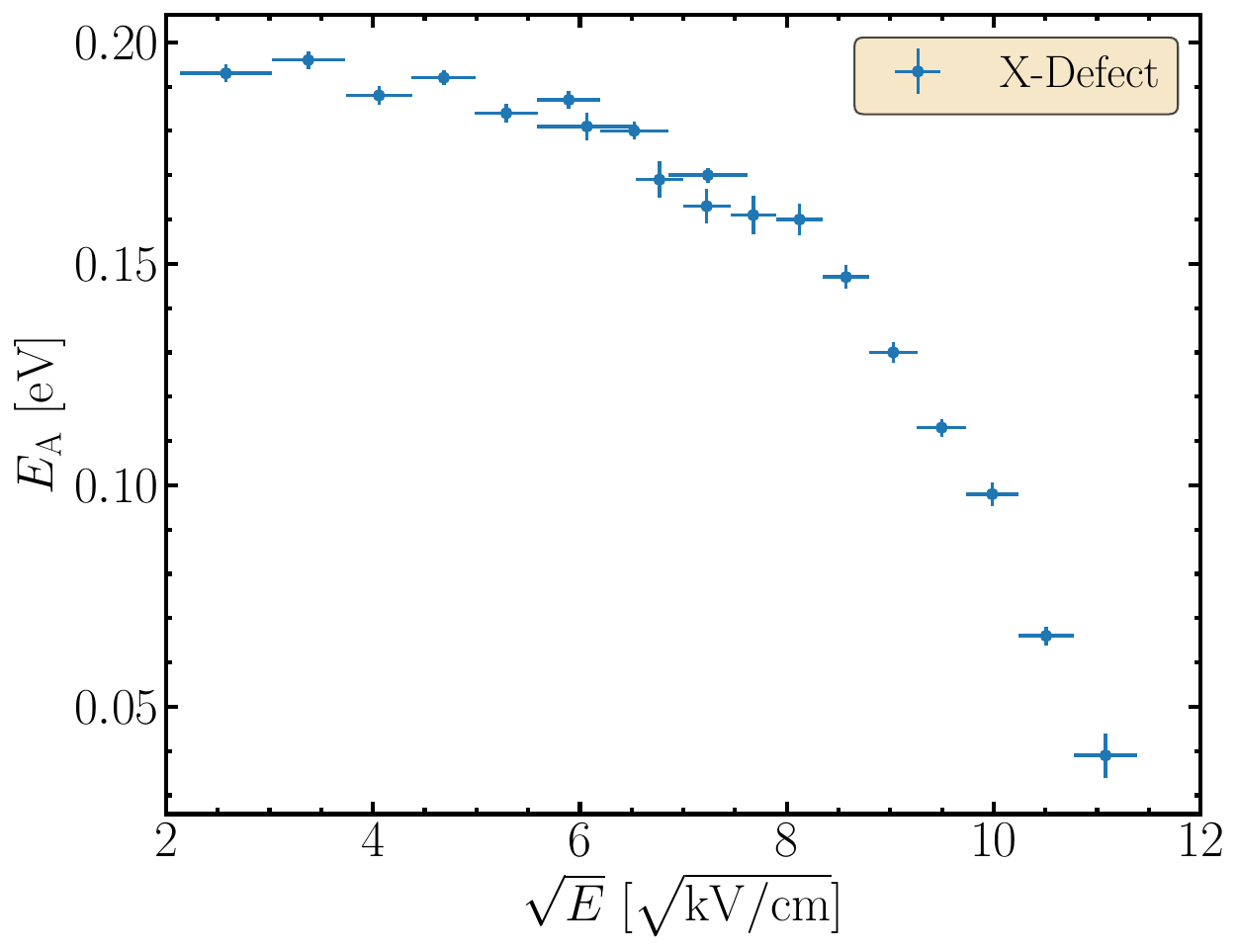}
    \caption{Re-creation of \autoref{fig:DLTS_efield_dependence}, but with the x-axis scaled to $\sqrt{E}$. A linear dependence would be expected for defects subject to the PF mechanism.}
    \label{fig:sqrtE}
\end{figure}
\begin{figure}[t]
    \centering
    \includegraphics[width=\linewidth]{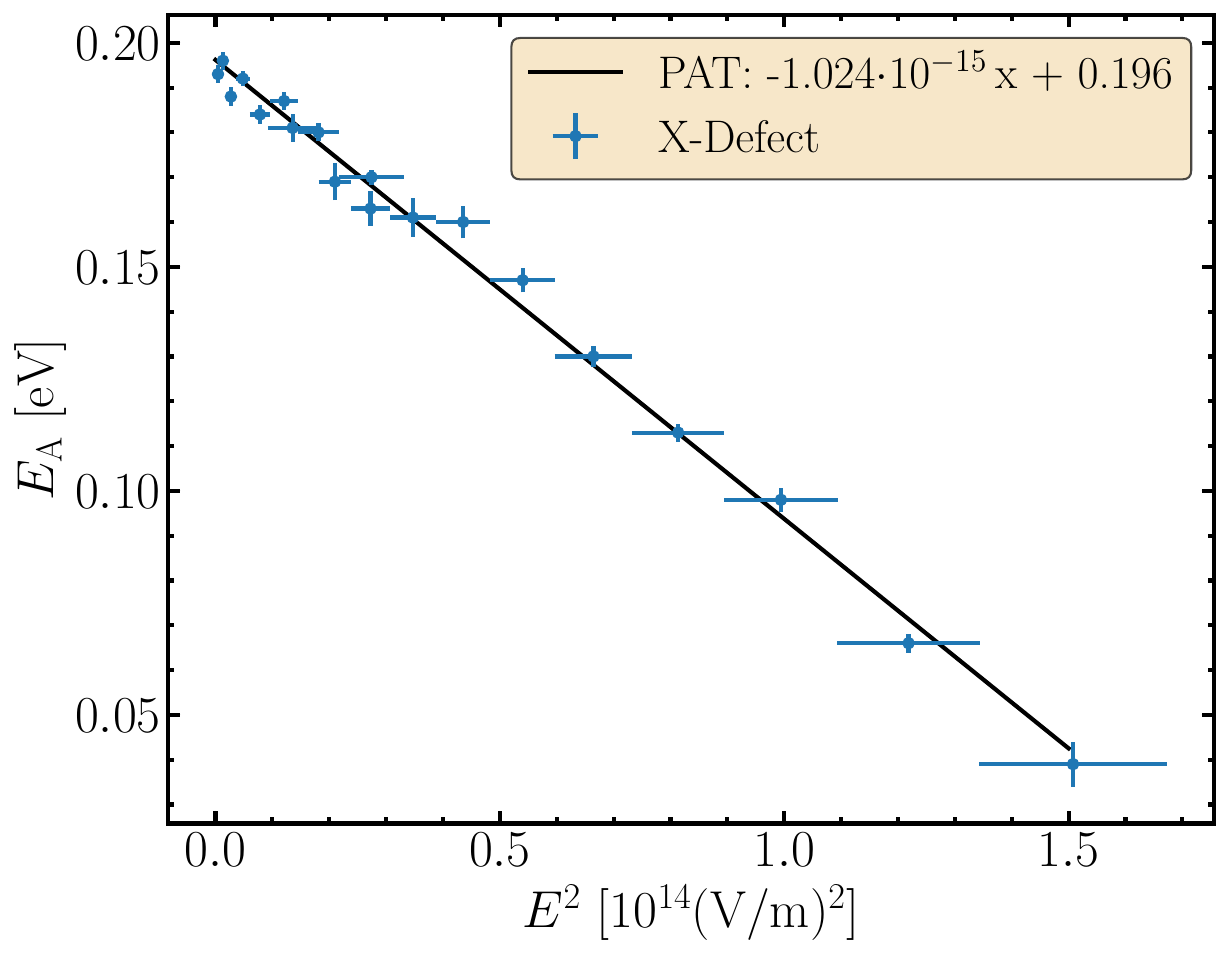}
    \caption{Re-creation of \autoref{fig:DLTS_efield_dependence}, but with the x-axis scaled to $E^2$. A linear dependence proofs the defect is subject to the PAT mechanism.}
    \label{fig:E2}
\end{figure}

For a better visualization of the field dependence, the activation energy data of the \Xdefect from \autoref{fig:DLTS_efield_dependence} is plotted in \autoref{fig:sqrtE} as a function of $\sqrt{E}$ and in \autoref{fig:E2} as a function of $E^2$. 
Evidently, the $E^2$ dependence results in a straight line as expected for the phonon-assisted tunnelling process.

In \autoref{fig:DLTS_efield_dependence} it looks like a small change in the activation energy is present for the \CiOi defect.
However, this is not expected and the observed dependence lies well within the present uncertainties, making a distinction between PF, PAT or a linear dependence not possible.
Hence, no conclusions can be drawn about this.

In conclusion, the comparison and good match of all measured characteristics of the \Xdefect energy level with the properties of the di-vacancy donor energy level, leads to the conclusion that the \Xdefect is indeed the Silicon di-vacancy \Vtwo.

\subsection{Simulation of Phonon-assisted Tunnelling}\label{sec:simulation}
To further test the conclusion of the \Xdefect being the \Vtwo, hence being subject to the PAT mechanism, TSC measurements of the \Xdefect under different bias conditions are compared to simulations using defect parameters obtained from DLTS measurements.
In the following, first a description of the PAT model used in the simulations is given, and then results from the simulations are compared to the measurements.

Phonon-assisted tunnelling causes a change of the emission rate $e$, exponentially with the squared electric field strength $E$ 
\begin{align}
    \Gamma_\text{PAT}(E,T)=\dfrac{e(E)}{e(0)}=\exp{\left(\dfrac{E^2}{E_\text{c}^2}\right)},\label{eq:emission}
\end{align}
with $\Gamma_\text{PAT}(E,T)$ being the field enhancement factor \cite{ganichev}.
Assuming Shockley-Read-Hall emission statistics \cite{shockley_read,hall}, where
\begin{align}
    e(E)&\propto\exp{\left(-\dfrac{E_\text{A}^\text{eff}(E)}{k_\text{B}T}\right)},
\end{align}
\autoref{eq:emission} can be reformulated to
\begin{align}
    E_\text{A}^\text{eff}(E)&=E_\text{A}(0)-k_\text{B}T\left(\dfrac{E}{E_\text{c}}\right)^2, \label{eq:PAT_E_lowering}
\end{align}
where $k_\text{B}$ is the Boltzmann constant, $T$ the temperature, $E$ the electric field strength and $E_\text{c}$ the so-called characteristic field strength
\begin{align}
    E_\text{c}&=\sqrt{\dfrac{3m^*\hbar}{q^2\tau_2^3}},\label{eq:Ec}
\end{align}
with $\hbar$ being the reduced Planck constant, $q$ the electric charge, $\tau_2$ the tunnelling time and $m^*$ the effective mass of the charge carriers \cite{ganichev, ganichev2}.
\autoref{eq:PAT_E_lowering} describes the behaviour seen in \autoref{fig:DLTS_efield_dependence} with \autoref{eq:E2}.
Combining this together with \autoref{eq:Ec} yields
\begin{align}
    \tau_2(T)=\left(\dfrac{-3m^*\hbar a}{q^2k_\text{B}T}\right)^{1/3}.
\end{align}
These relations can be used to simulate a PAT-enhanced emission process.
\begin{figure}[t]
    \centering
    \includegraphics[width=\linewidth]{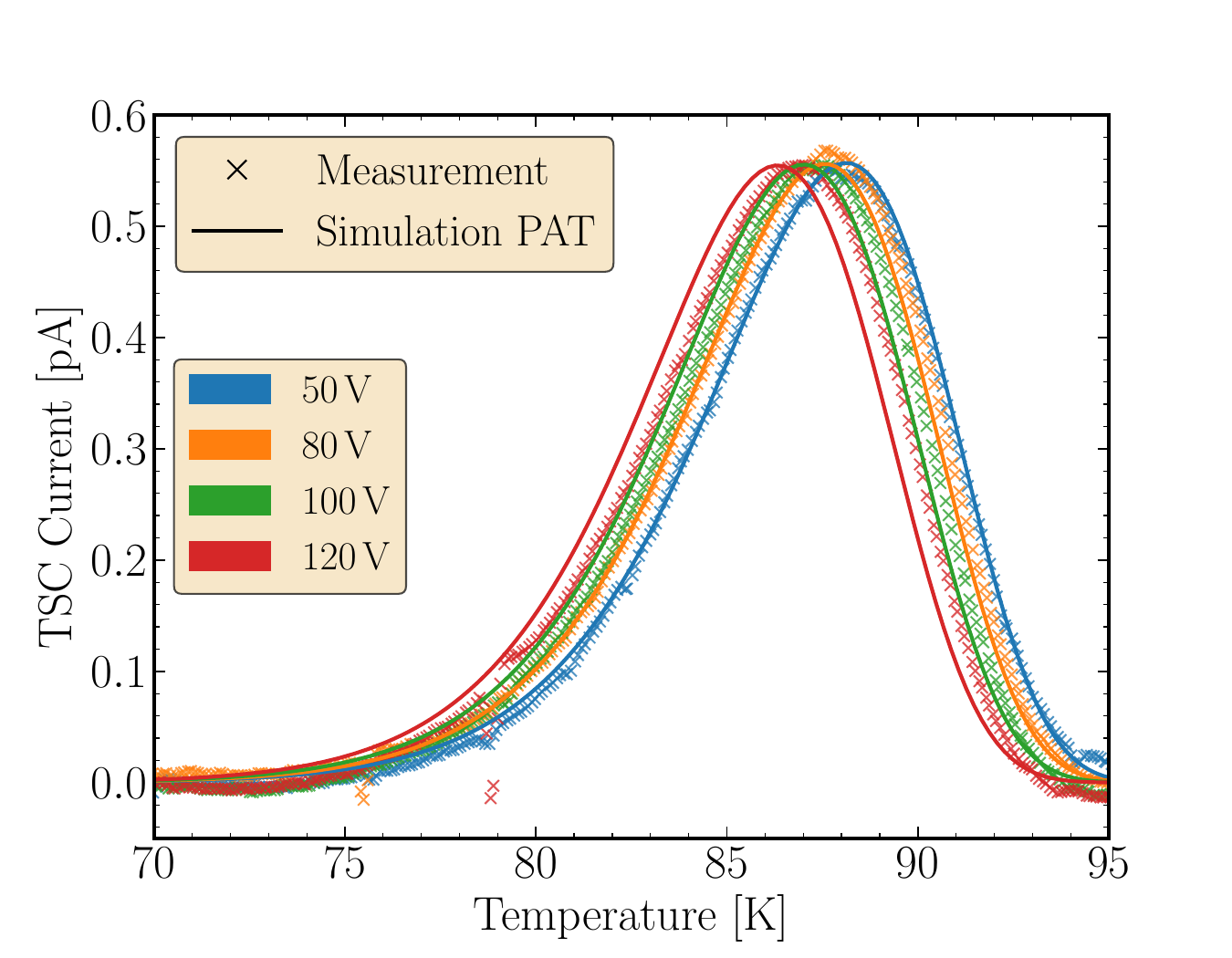}
    \caption{Comparison of TSC measurements on the $\gamma$-irradiated diode for different reverse biases with simulated spectra of the \Xdefect using the PAT model.}
    \label{fig:sim_gamma_PAT}
\end{figure}
\begin{figure}[t]
    \centering
    \includegraphics[width=\linewidth]{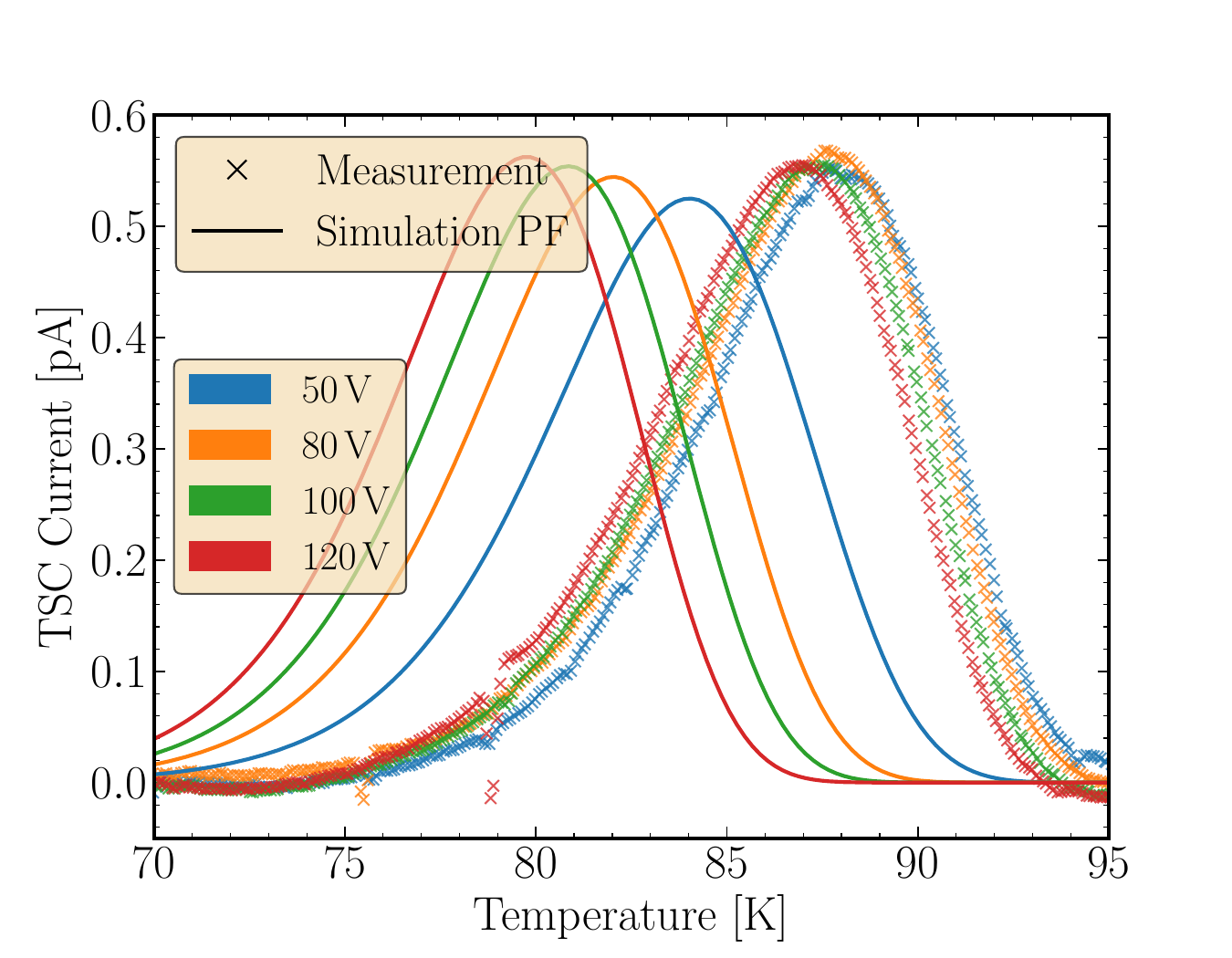}
    \caption{Comparison of TSC measurements on the $\gamma$-irradiated diode for different reverse biases with simulated spectra of the \Xdefect using the PF model.}
    \label{fig:sim_gamma_PF}
\end{figure}

The used simulation software is a fully custom python framework.
It is similar to the pyTSC framework used in References~\cite{anja2,niels_4HSiC}.
It follows the basic approach to model the thermally stimulated current using the Shockley-Read-Hall emission statistics, with activation energy $E_\text{A}$ and capture cross-section $\sigma$ as input parameters
\begin{align}
    e(T) = \sigma \cdot v_\text{th}(T) \cdot N_\text{c}(T) \cdot \exp\left(-\dfrac{E_\text{A}}{k_\text{B} T}\right).
\end{align}
The thermal velocity $v_\text{th}$ and density of states $N_\text{c}$ are calculated at each temperature step, where the temperature itself is a function of time $T(t)$.
The sensor thickness $d$ is divided into $n=\num{1000}$ slices, ensuring a sufficiently uniform electric field within each slice.
The emission term is then calculated for every temperature step in every slice, while the initialised trap occupancy is updated according to
\begin{align}
    n_\text{t}(T)=N_\text{t}(T_0)\exp{\left(-\dfrac{1}{\beta}\int\limits_{T_0}^{T}e(T)dT\right)},\label{eq:fractional_occupation_change}
\end{align}
with $\beta=\SI{11}{\kelvin\per\minute}$ being the heating rate.
The electric field profile and depletion depth is calculated according to \equationsref{eq:efield}, \ref{eq:efield_overdepletion} and \ref{eq:depletion_depth}.
\Neff is calculated dynamically for every temperature step and slice, accounting for the emission of trapped charges from defects providing space charge.
The emission term is modified by the weighting field for each slice, which in the case of a planar diode corresponds to the inverse of the depletion depth \cite{sze}.
The given defect concentration is assumed to be homogeneously distributed throughout the sensor bulk, the fraction inside the depleted volume is initialised as fully filled and the remainder as empty.
The current is determined by the product of the defect concentration $N_\text{t}$ times the fractional change of the trap occupancy, and summed over all slices and time steps. 
Taking the product of this with the electric charge $q$ and depleted volume results in a current with the unit of ampere.
The full equation becomes
\begin{align}{\scriptstyle
    I_\text{TSC}(T)=\dfrac{qA}{2} \sum\limits_{i=1}^{n} \sum\limits_\text{Defects} \left[ e(T) n_\text{t}(T) \Gamma(E,T) \right] \Delta z_i},
\end{align}
where $\Delta z_i$ are the $n$ slices of the sensor thickness and $n_\text{t}$ is the fraction of defect states occupied, which evolves with an exponential decay over time according to \autoref{eq:fractional_occupation_change}.
The $\Gamma$ factor comes from the different field enhancement models, which can be activated for each defect individually.
For defects with no field enhancement, $\Gamma$ is set to 1.
The PAT model is implemented as described by \autoref{eq:emission}, with the factor $\Gamma_\text{PAT}(E,T)$ being calculated for each slice and time step.
The PF model is implemented with $\Gamma_\text{PF}(E,T)$ to modify the emission term.
It is calculated with the three-dimensional description of the Poole-Frenkel effect from Hartke \cite{hartke} for each slice and time step
\begin{align}
    \Gamma_\text{PF}(E,T)=\dfrac{1}{\gamma^2}\left[e^\gamma(\gamma-1)+1\right]+\dfrac{1}{2},\label{eq:gamma_PF}
\end{align}
with
\begin{align}
    \gamma = \dfrac{q}{k_\text{B}T}\sqrt{\dfrac{qE}{\pi\varepsilon_0\varepsilon_\text{r}}}.
\end{align}
To validate the implementation of the PF model, a simulation of TSC spectra for the \BiOi defect was performed and compared to measurements.
This is shown and further discussed in \ref{sec:PF_validation}.

TSC measurements of the $\gamma$-irradiated diode for different bias voltages are displayed in \autoref{fig:sim_gamma_PAT} and \ref{fig:sim_gamma_PF} (crosses).
A shift of the \Xdefect's peak to lower temperatures with increasing bias is clearly visible.
Superimposed on these measurements are the simulated spectra (solid lines), using the zero-field activation energy from \autoref{fig:DLTS_efield_dependence} and the capture cross-section obtained from DLTS measurements on the same sample \cite{anja2}.
For this simulation the PAT enhancement model is used.
A good, but not perfect match between measurement and simulation is found.
At the lower bias voltages, nearly perfect matches are found.
At higher biases, i.e. stronger electric fields, the rising edge of the simulated peak is faster compared to the measured one, while the falling edge is not shifted sufficiently to lower temperatures.
For all but the highest biases, a good match of the positions of the peak maxima is observed.

\autoref{fig:sim_gamma_PF} shows what the effect of the PF would look like for the \Xdefect.
Clearly, this does not match the measured spectra.
The simulated TSC curves assuming a PF effect show peaks at lower temperatures than the measured data. 
This is in line with the experimental and fitted data shown in \autoref{fig:DLTS_efield_dependence}, where the activation energies from the PF fit are below those of the measurement at the relevant field strengths of the TSC measurements shown in \autoref{fig:sim_gamma_PF} (\num{14} to \SI{22}{\kilo\volt\per\centi\meter}).
The observed shift in the measurements do not arise from a PF mechanism.

An overall good agreement between the PAT model and the measurement data is obtained, even though the peak shapes could not be perfectly re-created by the simulation.
Additionally, with \autoref{fig:sim_gamma_PF} it can be clearly ruled out that a PF mechanism prevails.
In conclusion, these simulations further support the assignment of the \Xdefect to the \Vtwo.

\section{Conclusions}
A comprehensive experimental and simulation-based study on identifying the origin of the so-called \Xdefect in irradiated $p$-type silicon is presented in this work.
This defect is a hole trap and appears as a low-temperature shoulder to the \BiOi defect peak in TSC measurements. 
Although it was speculated in previous works that the defect could be the singly charged donor state of the di-vacancy \Vtwo, it was not identified as such. 
The observed field strength dependent emission rate was seemingly contradicting the interpretation as the \Vtwo should not be subject to a Poole-Frenkel effect. 

This misinterpretation is clarified in this work. 
By utilising a custom simulation framework it was possible to better link DLTS and TSC measurements in using DLTS-derived parameters to simulate TSC measurements. 
Furthermore, Difference-DLTS measurements were performed to study the dependence of the defect's activation energy on the electric field strength.
A clear quadratic dependence was found, providing apparent evidence that the phonon-assisted tunnelling (PAT) mechanism prevails rather than the Poole-Frenkel (PF) effect. 
This finding is in-line with the previous observation that the \Xdefect is not contributing to the effective doping concentration.
The identification of PAT as an emission mechanism implies that the \Xdefect is electrically neutral in the space charge region, and therefore does not affect \Neff at ambient temperatures.
Furthermore, the measured characteristics of the \Xdefect, including apparent activation energy, capture cross-section and emission mechanism, are in good agreement with literature values for the donor charge state of the singly charged di-vacancy, \Vtwo.
The zero-field activation energy of the \Xdefect is determined to be $E_\text{A,0}=\SI{0.1962(7)}{\electronvolt}$.

In conclusion, the defect previously introduced as the \Xdefect is indeed the singly charged Silicon di-vacancy, \Vtwo.

\FloatBarrier
\appendix

\section{Validation of the PF model}\label{sec:PF_validation}
\begin{figure}[t]
    \centering
    \includegraphics[width=\linewidth]{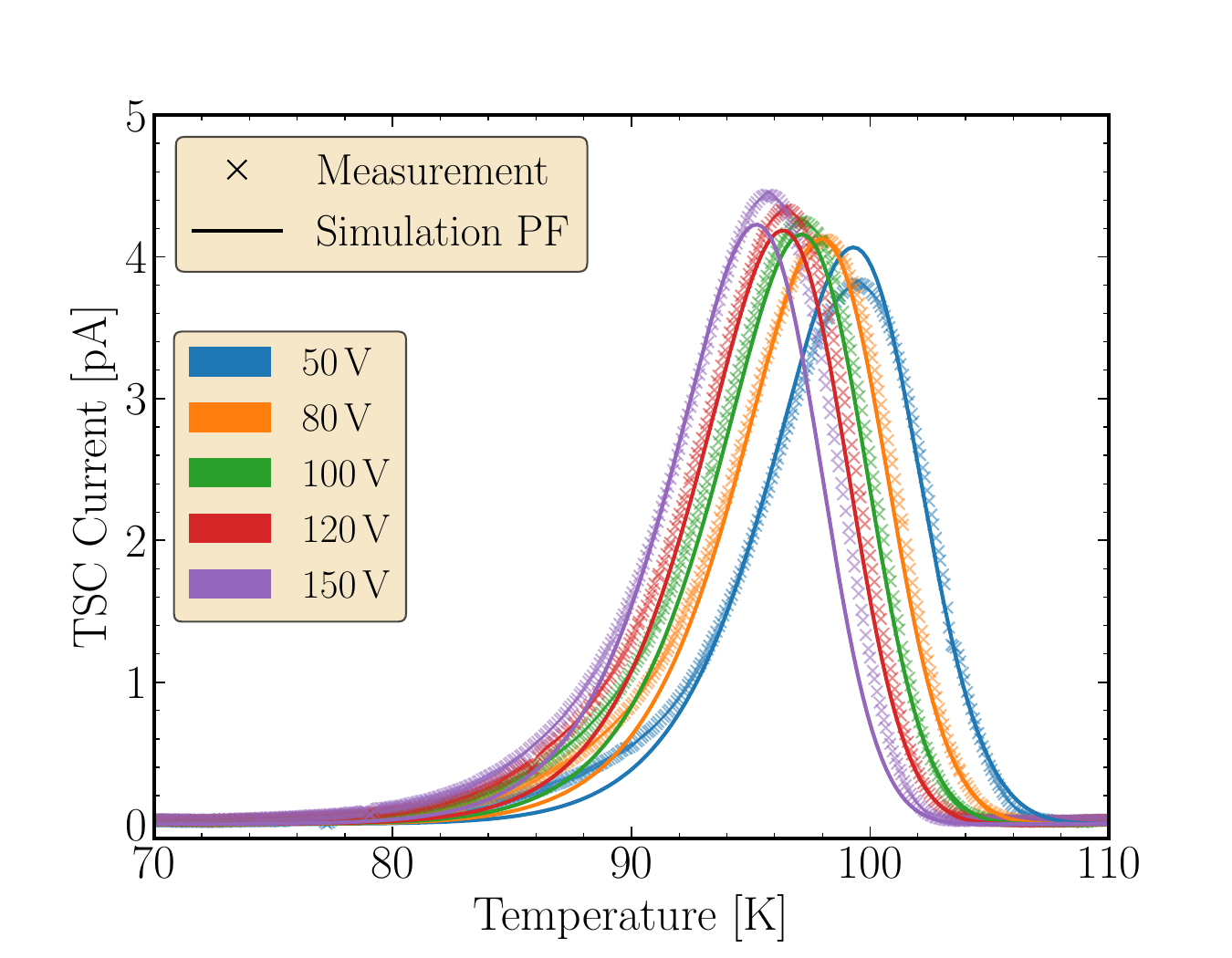}
    \caption{TSC measurements of the \BiOi defect at different bias voltages from the $\gamma$-irradiated diode, superimposed with the simulated spectra.}
    \label{fig:sim_BiOi_validation}
\end{figure}
\begin{figure}[t]
    \centering
    \includegraphics[width=\linewidth]{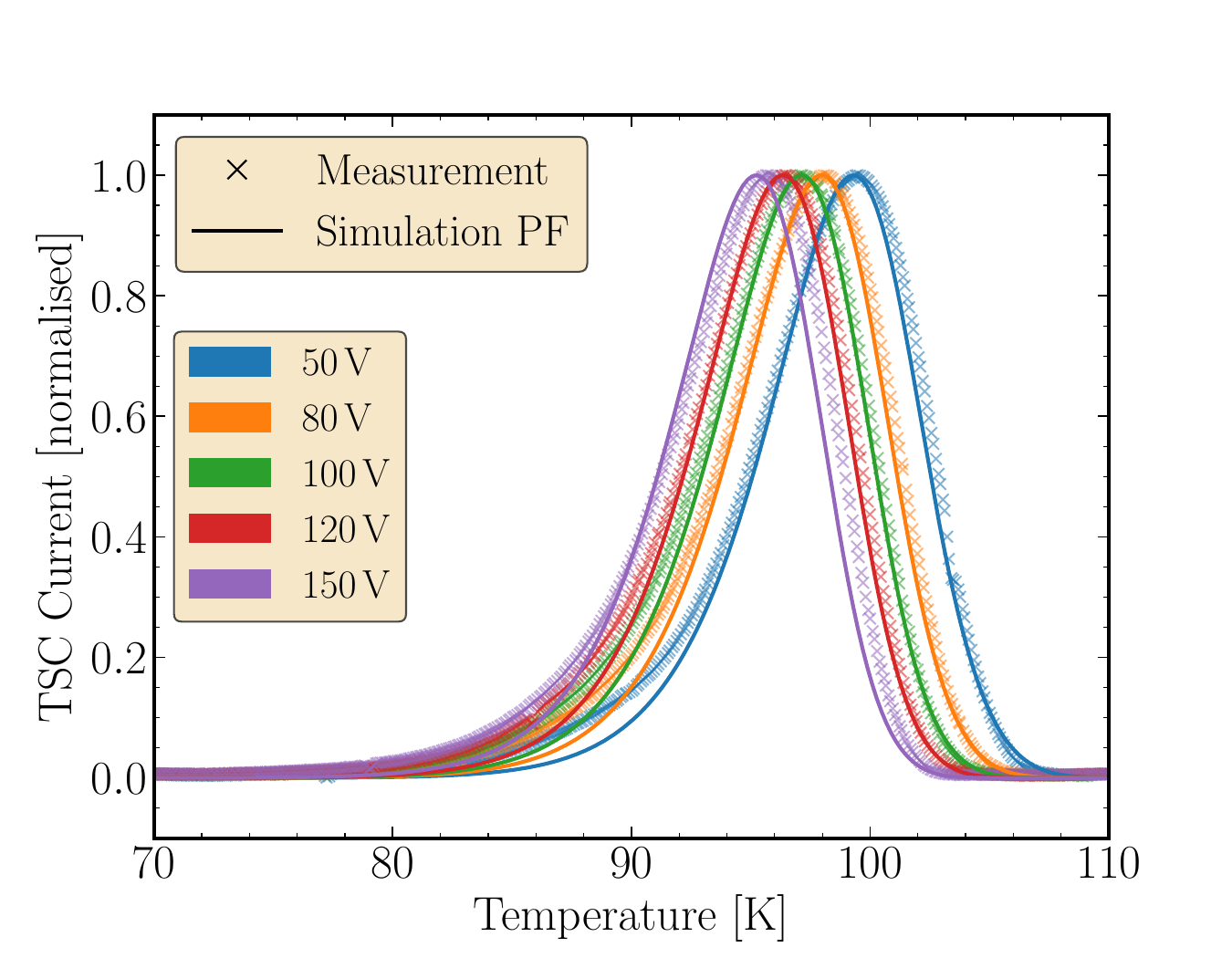}
    \caption{\autoref{fig:sim_BiOi_validation}, but with the currents normalised to each peak's amplitude.}
    \label{fig:sim_BiOi_validation_normalised}
\end{figure}
To validate the PF model implementation in the simulation code, TSC measurements for the \BiOi defect were carried out on the $\gamma$-irradiated diode at different bias voltages, as the \BiOi defect is known to be subject to the PF effect.
The measurements alongside the simulation are shown in \autoref{fig:sim_BiOi_validation} and \ref{fig:sim_BiOi_validation_normalised}.
An activation energy of \SI{0.27}{\electronvolt} and a capture-cross section of \SI{1.05e-14}{\centi\meter\squared} were used as input parameters, taken from Ref.~\cite{liao1}.
To obtain the simulated spectra shown in \autoref{fig:sim_BiOi_validation}, the emission enhancement factor $\Gamma_\text{PF}$ had to be modified by a factor of \num{0.97}, a modification of \SI{3}{\percent}.
This adjustment is justified by the requirement formulated in Ref.~\cite{hartke}.
Hartke states that the PF effect arises from the electrostatic attraction between a singly charged positive ion and an electron under the influence of a uniform electric field.
The requirement of a uniform field might not be satisfied for a highly doped, irradiated diode.
Therefore, the strength of this attraction is modified.
Similar corrections are reported elsewhere \cite{liao2}.
With this modification, the model shows good agreement with the experimental data.



\section*{Acknowledgements}
The work was performed partly in the framework of the CERN RD50 and DRD3 collaborations.

Niels Sorgenfrei acknowledges that his work has been sponsored by the Wolfgang Gentner Programme of the German Federal Ministry of Research, Technology and Space (grant no. 13E18CHA).

Ioana Pintilie acknowledges the funding received through the IFA-CERN-RO 07/2024 project.

\section*{Declaration of competing interest}
The authors declare that they have no known competing financial interests or personal relationships that could have influenced the work reported in this paper.

\section*{Data availability}
The data will be made available on request.

\FloatBarrier
\bibliographystyle{elsarticle-num} 
\bibliography{literature}

\end{document}